# Construction of weakly CUD sequences for MCMC sampling

## Seth D. Tribble[*]

*Morgan-Stanley Inc.*
e-mail: stribbs@gmail.com

## Art B. Owen[*]

*Stanford University*
e-mail: owen@stat.stanford.edu

**Abstract:** In Markov chain Monte Carlo (MCMC) sampling considerable thought goes into constructing random transitions. But those transitions are almost always driven by a simulated IID sequence. Recently it has been shown that replacing an IID sequence by a weakly completely uniformly distributed (WCUD) sequence leads to consistent estimation in finite state spaces. Unfortunately, few WCUD sequences are known. This paper gives general methods for proving that a sequence is WCUD, shows that some specific sequences are WCUD, and shows that certain operations on WCUD sequences yield new WCUD sequences. A numerical example on a 42 dimensional continuous Gibbs sampler found that some WCUD inputs sequences produced variance reductions ranging from tens to hundreds for posterior means of the parameters, compared to IID inputs.

**AMS 2000 subject classifications:** Primary 62F15; secondary 11K45, 11K41.

**Keywords and phrases:** completely uniformly distributed, Gibbs sampler, equidistribution, probit, quasi-Monte Carlo.

## 1. Introduction

In Markov chain Monte Carlo (MCMC) sampling, random inputs are used to determine a sequence of proposals and, in some versions, their acceptance or rejection. Considerable effort and creativity have been applied to devising good proposal mechanisms. Many examples can be found in recent books such as Newman and Barkema (1999), Liu (2001), Robert and Casella (2004), and Landau and Binder (2005).

With very few exceptions, described below, the proposals and acceptance-rejection decisions are all sampled the same way. A sequence of points $u_i \in (0,1)$ that simulate independent draws from the $U[0,1]$ distribution is used to drive the process. By replacing IID draws with more balanced samples, we can hope to improve the accuracy of MCMC.

[*]Supported by grants DMS-0306612 and DMS-0604939 from the U.S. NSF. We thank Richard Simard for answering questions about SSJ and an anonymous reviewer for helpful comments.

MCMC sampling is subtle and modifying the IID driving sequence without theoretical support is risky. To do so is to simulate a Markov chain using random variables that do not have the Markov property. Caution is in order and few have tried it. According to Charles Geyer, writing in 2003, (<http://www.stat.umn.edu/geyer/mcmc/talk/mcmc.pdf>) "Every MCMC-like method is either a special case of the Metropolis-Hastings-Green algorithm, or is bogus". Our objective in this paper is to show that a variety of alternative sampling schemes avoid the latter category. They yield consistent MCMC estimates whose accuracy we investigate empirically.

Two exceptions to IID driving sequences that we know of are Liao (1998) and Chaudary (2004). The first proposed using randomly reordered quasi-Monte Carlo points in the Gibbs sampler. The second proposed strategic sampling of proposals with weighting of rejected proposals. Both found empirical evidence of a modest improvement, but neither gave any theory to support his method.

We will follow up on Liao's proposal, which is illustrated in Figure 1. The quasi-Monte Carlo points, which are rows of the matrix $A$ are very uniformly distributed through $[0,1]^s$. They get reordered into a matrix $X$ that then has its rows concatenated into a long vector $u$ that gets used to drive the MCMC computation. We refer to both $u$ and $X$ as the driving sequence.

Recently Owen and Tribble (2005) proved consistency for Metropolis-Hastings algorithms, including Gibbs sampling as a special case, when the driving sequence $u_i$ are completely uniformly distributed (CUD), or even weakly CUD (WCUD), as defined below. That work built on Chentsov (1967) who gave conditions under which CUD sequences lead to consistency when Markov chains are sampled by inversion.

We give formal definitions of CUD and WCUD sequences below. For now, note that a CUD sequence is one that can be broken into overlapping $d$-tuples that pass a uniformity test, and that this holds for all $d$. Replacing IID points by CUD points is similar to using the entire period of a random number generator. When the points of the CUD sequence are well balanced, the effect is to get a quasi-Monte Carlo (QMC) version of MCMC. One then has to design or choose random number generators with good uniformity that are small enough to use in their entirety.

The theory in Owen and Tribble (2005) applied to infinite sequences, but the examples there used finite sequences that were not yet known to be WCUD. This paper establishes that several classes of constructions give WCUD sequences and shows how certain natural operations on WCUD sequences yield other WCUD sequences. We then illustrate the use of WCUD sampling on a 42 parameter Gibbs sampling problem and find that the posterior means are estimated with variance reductions ranging from tens to hundreds. A detailed outline of this paper follows.

Section 2 provides background material on MCMC, QMC, and CUD sequences. In Section 3 a new definition of triangular array (W)CUD sequences is made, suitable for QMC constructions that are not initial segments of infinite sequences. Theorem 2 shows that triangular array (W)CUD sequences lead to consistent MCMC estimates. Sometimes it is simpler to prove a pointwise version of the (W)CUD property instead of the uniform one required. Lemma 1 shows that pointwise (W)CUD sequences are necessarily (W)CUD. (W)CUD sequences are defined through overlapping blocks of consective values, but some QMC constructions make it easier to work with non-overlapping blocks. Theorem 3 shows that a (W)CUD like property defined over non-overlapping blocks suffices to prove the original (W)CUD property. That theorem also shows that one can restrict attention to any convenient infinite set of dimensions.

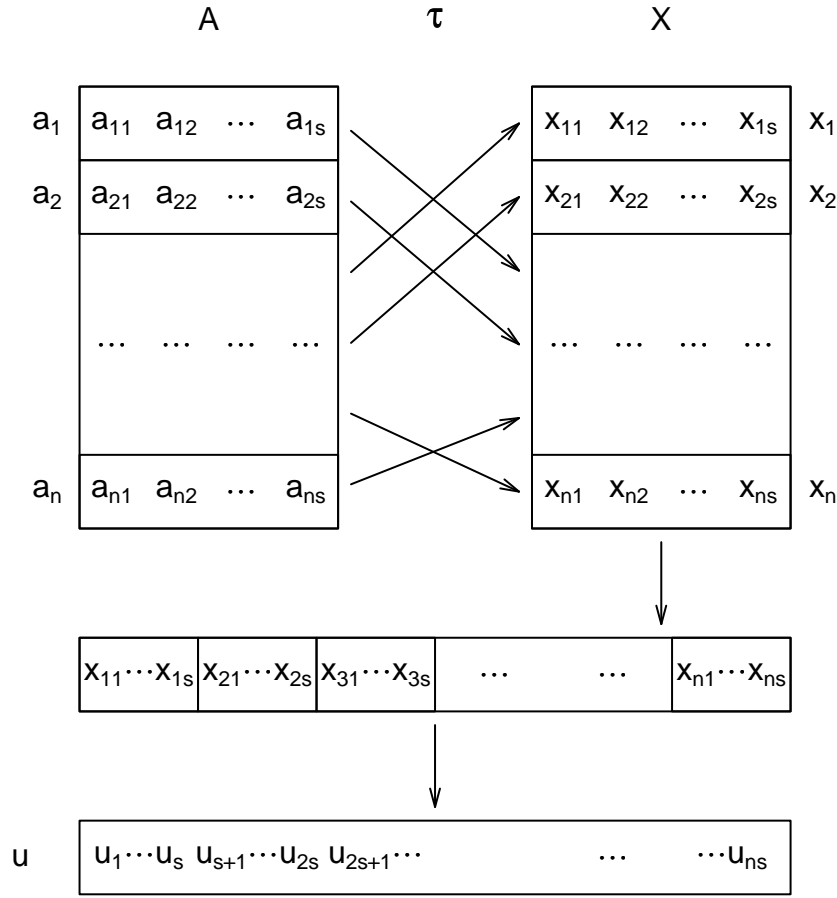

FIG 1. *This figure illustrates the construction used in Liao (1998). The matrix $A \in [0,1]^{n \times s}$ has its rows randomly reordered by a permutation $\tau$, creating the matrix $X \in [0,1]^{n \times s}$. The rows of $X$ are then concatenated into one long row vector whose contents are then used to define the vector $u \in [0,1]^{ns}$. The values of $u$ are then fed into the MCMC algorithm.*

Next we turn to specific constructions. In Section 4, Theorem 4 shows that a lattice construction of Niederreiter (1977) leads to a CUD sequence. Cranley-Patterson rotations are commonly used to randomize lattice rules. Lemma 2 shows that Cranley-Patterson rotations of Niederreiter's sequence, or indeed of any (W)CUD sequence, are WCUD. Section 5 investigates Liao's proposal for randomly reordering the points of a QMC sequence. Lemma 3 gives a non-asymptotic bound for the discrepancy of the reordered sequence in terms of the discrepancy of the original one. Then Theorem 5 gives conditions for the reordered points to be WCUD. Liao's proposal simply shuffles a QMC sequence. Theorem 6 shows that some generalizations that rearrange the QMC points are also WCUD. Section 6 shows that we can mix IID $U[0,1]$ sequences into WCUD sequences in a certain way, and end up with a WCUD sequence. This then proves, via a coupling argument, that Liao's proposal for handling acceptance-rejection sampling leads to consistent estimates.

Section 7 presents an example, with a probit model on some data of Finney (1947), using QMC-MCMC to drive the Gibbs sampler from Albert and Chib (1993). In this example, the parameter vector has 42 dimensions. The variance reductions attained are typically in the range from 20 to 40 but some improvements of over 500-fold were attained. Section 8 summarizes our findings, discusses rates of convergence and extensions to continuous state spaces.

To conclude this introduction, we mention some related work in the literature that merges QMC and MCMC ideas. L'Ecuyer et al. (2005) have developed an array-RQMC method for simulating Markov chains on an ordered state space. Craiu and Lemieux (2007) combine antithetic and stratified sampling in the multiple-try Metropolis algorithm of Liu et al. (2000). James Propp and co-workers have been applying QMC ideas to derandomize some randomized automata. At present the best way to find this work is via Internet search using the term "rotor-router". Lemieux and Sidorsky (2005) use quasi-Monte Carlo sampling to drive the exact sampling of Propp and Wilson (1996).

## 2. Background

This section sets out the notation for MCMC, assuming some familiarity with Markov chain Monte Carlo methods, as described for example in Liu (2001), Robert and Casella (2004) or Gilks et al. (1996). Our version works on a finite state space with Metropolis-Hastings type proposals.

Then we describe quasi-Monte Carlo sampling (QMC) and use it to define CUD and weakly CUD sequences. Finally we show how CUD and weakly CUD sequences can be used as driving sequences for Metropolis-Hastings sampling.

We use integers $m$, $s$, and $d$ to describe dimensions in this paper. One Metropolis-Hastings step typically consumes $m$ uniform random numbers. A quasi-Monte Carlo point set is typically constructed for a finite dimension $s$. It is most natural to arrange matters so that $m = s$, but we need two distinct integers because $m \neq s$ is also workable. Finally we will need some equidistribution properties to hold for all $d$ in an infinite set of dimensions.

Points in $[0,1]^m$, $[0,1]^s$, or $[0,1]^d$ are denoted by letters such as $x$ and $z$, or when there are many such points, by $x^{(i)}$ or $z^{(k)}$ for integer indices. Components of such tuples are denoted as $x_j$ or $z_j^{(k)}$. For $j \leq k$, the notation $x_{j:k}$ denotes the $(k-j+1)$–tuple taken from components $j$ through $k$ inclusive of $x$.

### 2.1. Markov chain Monte Carlo

We describe MCMC for sampling from a distribution $\pi$ on a discrete state space $\Omega = \{\omega_1, \ldots, \omega_K\}$ for $K < \infty$. We suppose as is usual that the ratio $\pi(\omega)/\pi(\omega')$ can be easily obtained for any pair $\omega, \omega' \in \Omega$, or at least for any pair that can arise as two consecutive samples. Starting with some point $\omega^{(0)} \in \Omega$, the MCMC simulation generates a Markov chain $\omega^{(i)} \in \Omega$ for $i \geq 1$ whose stationary distribution is $\pi$.

Common usage of Metropolis-Hastings sampling make a proposal $\widetilde{\omega}^{(i+1)}$ based on $\widetilde{\omega}^{(i)}$ and $m-1$ consecutive $u_j$ from the driving sequence:

$$\widetilde{\omega}^{(i+1)} = \Psi(\omega^{(i)}, u_{mi+1}, \ldots, u_{mi+m-1}). \tag{1}$$

Acceptance or rejection of the proposal is based on one more member of the driving sequence as follows:

$$\omega^{(i+1)} = \begin{cases} \widetilde{\omega}^{(i+1)}, & u_{mi+m} \leq A(\omega^{(i)} \to \widetilde{\omega}^{(i+1)}) \\ \omega^{(i)}, & \text{else,} \end{cases} \tag{2}$$

using the Metropolis-Hastings acceptance probability

$$A(\omega \to \widetilde{\omega}) = \min\left(1, \frac{\pi(\widetilde{\omega})p(\widetilde{\omega} \to \omega)}{\pi(\omega)p(\omega \to \widetilde{\omega})}\right). \tag{3}$$

Here $p(\omega \to \widetilde{\omega})$ denotes the probability of proposing a transition from $\omega$ to $\widetilde{\omega}$.

The mechanism described by equations (1), (2), and (3) is less general than the usual MCMC. It leaves out the case where acceptance-rejection sampling is used to generate one or more of the components of the proposal. Section 6 shows how one can splice in IID elements of the driving sequence for that case.

The function $\Psi$ that constructs proposals may involve inversion of CDFs or apply other similar transformations for generating random variables given by Devroye (1986). We will however need a regularity condition (Jordan measurability) for the proposals. This very mild condition rules out pathological constructions.

**Definition 1** (Regular proposals). *The proposals are regular if for all $\omega, \widetilde{\omega} \in \Omega$ the set*

$$S_{\omega \to \widetilde{\omega}} = \{(u_1, \ldots, u_{m-1}) \mid \widetilde{\omega} = \Psi(\omega, u_1, \ldots, u_{m-1})\} \subseteq [0,1]^{m-1}$$

*is Jordan measurable.*

Jordan measurability of $S_{\omega \to \tilde{\omega}}$ means that it has a Riemann integrable indicator function. The boundary of such sets has volume zero. Proposals that do one thing for $u_i$ with rational components and another for irrational $u_i$ might be ruled out. Practically useful proposal mechanisms are regular.

By stage $n$ the fraction of time spent in state $\omega$ is

$$\hat{\pi}_n(\omega) = \frac{1}{n} \sum_{i=1}^{n} 1_{\omega^{(i)}=\omega}. \tag{4}$$

Consistency of $\hat{\pi}_n$ is defined differently for random and nonrandom $u_i$, as follows.

**Definition 2.** *The chain is consistent if*

$$\lim_{n \to \infty} \hat{\pi}_n(\omega) = \pi(\omega) \tag{5}$$

*holds for all $\omega, \omega^{(0)} \in \Omega$.*

**Definition 3.** *The chain is weakly consistent if*

$$\lim_{n \to \infty} \Pr(|\hat{\pi}_n(\omega) - \pi(\omega)| > \epsilon) = 0 \tag{6}$$

*holds for all $\omega, \omega^{(0)} \in \Omega$ and for all $\epsilon > 0$.*

### 2.2. Quasi-Monte Carlo

In quasi-Monte Carlo sampling, one does not simulate randomness. Instead one picks points that are more uniform than random points would be. Here we sketch QMC sampling. The reader seeking more information may turn to Niederreiter (1992).

For a point $z = (z_1, \ldots, z_s) \in [0,1]^s$ let $[0, z]$ denote the $s$ dimensional box bounded by $(0, \ldots, 0)$ at the lower left and $z$ at the upper right. That is $[0, z]$ is the Cartesian product $\prod_{j=1}^{s}[0, z_j]$. The volume of this box is $\mathrm{Vol}([0, z]) = \prod_{j=1}^{s} z_j$. Given $n$ points $x^{(1)}, \ldots, x^{(n)} \in [0,1]^s$ define the empirical volume of the box as

$$\widehat{\mathrm{Vol}}_n([0, z]) = \frac{1}{n} \sum_{i=1}^{n} 1_{x^{(i)} \in [0, z]}$$

and the local discrepancy function

$$\delta_n^s(z) = \delta_n^s(z; x^{(1)}, \ldots, x^{(n)}) = \left| \widehat{\mathrm{Vol}}_n([0, z]) - \mathrm{Vol}([0, z]) \right|. \tag{7}$$

We suppress the dependence of $\widehat{\mathrm{Vol}}$ on $s$ and $x^{(i)}$ to keep notation uncluttered. The star discrepancy of $x^{(1)}, \ldots, x^{(n)} \in [0,1]^s$ is

$$D_n^{*s} = D_n^{*s}(x^{(1)}, \ldots, x^{(n)}) = \sup_{z \in [0,1]^s} \delta_n(z; x^{(1)}, \ldots, x^{(n)}). \tag{8}$$

The star discrepancy is an $s$ dimensional generalization of the Kolmogorov-Smirnov distance between $U[0,1]^s$ and the empirical distribution of the points $x^{(i)}$.

Let the function $f : [0,1]^s \to \mathbb{R}$ have total variation $\|f\|_{\text{HK}}$ in the sense of Hardy and Krause. For $s = 1$, $\|f\|_{\text{HK}}$ reduces to the usual concept of the total variation of a function. For $s > 1$ one must take care to use a measure of variation that does not vanish when $f$ depends only on a subset of its arguments. Variation in the sense of Hardy and Krause does so, as described in Owen (2005). Then the Koksma-Hlawka inequality

$$\left| \int_{[0,1]^s} f(x)\,dx - \frac{1}{n} \sum_{i=1}^n f(x^{(i)}) \right| \leq D_n^{*s} \times \|f\|_{\text{HK}} \qquad (9)$$

shows the advantage of low discrepancy for integration. There are QMC constructions for which $D_n^{*s} = O(n^{-1+\epsilon})$ holds for any $\epsilon > 0$. Then functions $f$ of finite variation can be integrated with an error rate that is superior to the familiar $O(n^{-1/2})$ root mean square error rate of Monte Carlo. The asymptotics can be slow to set in but in practice the accuracy of QMC ranges from comparable with MC to far superior to MC.

### 2.3. CUD and weakly CUD sequences

Quasi-Monte Carlo points will not always lead to the right answer when used to drive MCMC sampling. The QMC constructions that can be made to work are the ones that are completely uniformly distributed (CUD) and weakly CUD (WCUD) as defined below.

**Definition 4** (CUD). *The infinite sequence $u_i \in [0,1]$ for $i \geq 1$ is completely uniformly distributed, if*

$$\lim_{n \to \infty} D_n^{*d}\Big((u_1,\ldots,u_d), (u_2,\ldots,u_{d+1}), \ldots (u_n,\ldots,u_{d+n-1})\Big) = 0 \qquad (10)$$

*holds for every integer $d \geq 1$.*

Knuth (1998) describes several working definitions of randomness for random number generators. One of them is that the sequence be CUD. The concept of CUD sequences originated with Korobov (1948). Levin (1999) gives a recent survey of CUD constructions.

**Definition 5** (Weak CUD). *The infinite sequence of random variables $u_i \in [0,1]$ for $i \geq 1$ is weakly completely uniformly distributed, if*

$$\lim_{n \to \infty} \Pr\Big( D_n^{*d}\Big((u_1,\ldots,u_d), (u_2,\ldots,u_{d+1}), \ldots (u_n,\ldots,u_{d+n-1})\Big) > \epsilon \Big) = 0 \quad (11)$$

*holds for every integer $d \geq 1$ and every $\epsilon > 0$.*

**Theorem 1.** *Let $\omega^{(0)} \in \Omega = \{\omega_1, \ldots, \omega_K\}$. For $i \geq 0$ let $\widetilde{\omega}^{(i+1)}$ be a regular proposal generated from $(u_{mi+1}, \ldots, u_{mi+m-1})$ and $\omega^{(i)}$ via equation (1), and let* $\omega^{(i+1)}$ be determined from $u_{is}$ by equation (2). Assume that weak consistency (6) holds for all $\omega, \omega^{(0)} \in \Omega$ and all $\epsilon > 0$ when $u_i$ are independent $U[0,1]$ random variables. If the IID $u_i$ are replaced by weakly CUD $u_i$, then the weak consistency result (6) still holds. If the IID $u_i$ are replaced by CUD $u_i$, then the stronger consistency result (5) holds.

*Proof.* Owen and Tribble (2005), Theorem 3. □

Theorem 1 does not explicitly make any assumptions about whether the chain is ergodic or even whether it is irreducible. Those considerations are very important, but they are buried in the weak consistency assumption (6). If a proposal mechanism is known to be consistent for an IID driving sequence, then it remains so for driving sequences that are CUD or WCUD. Therefore MCMC sampling separates into two problems. One is selecting a good proposal mechanism. The other is selecting a weakly CUD driving sequence, where of course IID points constitute the usual choice.

## 3. Extensions of WCUD

This section presents some basic results for (W)CUD sequences that we use later to show that specific constructions give rise to consistent MCMC sampling. We define a triangular array notion of (W)CUD sequences for use with finite driving sequences.

Theorem 1, taken from Owen and Tribble (2005) applies to infinite sequences $u_i$. In practice one uses a finite sequence of length $N$. A theory for large $N$ has to account for the fact that the constructions are not always nested: for $N_1 < N_2$ the sequence of length $N_2$ might not be an extension of the sequence of $N_1$ points.

To formulate our limit, we take a triangular array in which the row indexed by $N$ has points $u_{N,1}, \ldots, u_{N,N} \in [0,1]$ that we use to compute $\hat{\pi}$. The value $N$ belongs to an infinitely large nonrandom set $\mathcal{N}$ of positive integers. In terms of Figure 1, the bottom row has $N = ns$ numbers in $[0,1]$. Asymptotics for Liao's construction involve a limit of such figures indexed by $N$. Other constructions similarly involve limits of finite sequences.

The number $n$ of transitions made depends on $N$. When each transition consumes $m$ points of the driving sequence, then $n(N) = \lfloor N/m \rfloor$. The estimate using $u_{N,1}, \ldots, u_{N,N}$ is $\hat{\pi}_{n(N)}(\omega)$. Consistency or weak consistency holds when for all $\omega$, $\hat{\pi}_{n(N)}(\omega)$ converges, or converges weakly, to $\pi(\omega)$ as $N \to \infty$. Limits as $N \to \infty$ are always understood to be through the values $N \in \mathcal{N}$.

**Definition 6** (Triangular array (W)CUD). *Let $u_{N,1}, \ldots, u_{N,N} \in [0,1]$ for an infinite set of positive integers $N \in \mathcal{N}$. Suppose that as $N \to \infty$ through the values in $\mathcal{N}$, that*

$$D^{*d}_{N-d+1}((u_{N,1}, \ldots, u_{N,d}), \ldots, (u_{N,N-d+1}, \ldots, u_{N,N})) \to 0$$

*holds for any integer $d \geq 1$. Then the triangular array $(u_{N,i})$ is CUD. If the $u_{N,i}$ are random and*

$$\Pr(D^{*d}_{N-d+1}((u_{N,1},\ldots,u_{N,d}),\ldots,(u_{N,N-d+1},\ldots,u_{N,N})) > \epsilon) \to 0$$

*as $N \to \infty$ through values in $\mathcal{N}$ holds for all integers $d \geq 1$ and all $\epsilon > 0$, then the triangular array $(u_{N,i})$ is weakly CUD.*

If an infinite sequence $u_1, u_2, \ldots$ is CUD (or WCUD) then the triangular array of prefixes taking $u_{N,i} = u_i$, for all $N \geq 1$ is also CUD (respectively WCUD). This means that the triangular array definitions are broader than the original ones.

**Theorem 2.** *Suppose that the transitions are as described in Theorem 1, including weak consistency (6) when $u_i$ are independent $U[0,1]$ random variables. Let each transition consume $m$ of the $u_i$. If $u_i$ are replaced by elements $u_{N,i}$ of a CUD triangular array then*

$$\lim_{N\to\infty} \hat{\pi}_{\lfloor N/m \rfloor}(\omega) = \pi(\omega). \tag{12}$$

*If $u_i$ are replaced by elements $u_{N,i}$ of a WCUD triangular array then*

$$\lim_{N\to\infty} \Pr(|\hat{\pi}_{\lfloor N/m \rfloor}(\omega) - \pi(\omega)| > \epsilon) = 0. \tag{13}$$

*Proof.* The proof is similar to that of Theorem 3 in Owen and Tribble (2005), and so we only sketch it. Fix $\epsilon > 0$ and identify the set $\mathcal{T}_r(\epsilon) \subset [0,1]^{rm}$ for which $\sum_{\omega \in \Omega} \Pr(|\hat{\pi}_r(\omega) - \pi(\omega)| > \epsilon) > \epsilon$ holds when $(u_1, \ldots, u_{rm}) \in \mathcal{T}_r(\epsilon)$. For large enough $r$ the set $\mathcal{T}_r(\epsilon)$ has volume no more than $\epsilon$. Then apply Definition 6 using $d = rm$. The rest of the proof follows as in Owen and Tribble (2005). □

In proving that $D^{*d}_n$ converges to zero, the natural first step is to show that the local discrepancy $\delta^d_n(z)$ tends to zero at each $z$. It turns out that such a pointwise (W)CUD property implies the (W)CUD property.

**Definition 7** (Pointwise (W)CUD). *The triangular array $u_{N,i} \in [0,1]$ is pointwise CUD, if*

$$\lim_{N\to\infty} \delta^d_{N-d+1}\Big(z; (u_{N,1},\ldots,u_{N,d}),\ldots,(u_{N,N-d+1},\ldots,u_{N,N})\Big) = 0 \tag{14}$$

*holds for every integer $d \geq 1$ and every $z \in [0,1]^d$. The triangular array of random variables $u_{N,i} \in [0,1]$ is pointwise weakly CUD, if*

$$\lim_{N\to\infty} \Pr\Big(\delta^d_{N-d+1}\Big(z; (u_{N,1},\ldots,u_{N,d}),\ldots,(u_{N,N-d+1},\ldots,u_{N,N})\Big) > \epsilon\Big) = 0 \tag{15}$$

*holds for every integer $d \geq 1$, every $z \in [0,1]^d$, and every $\epsilon > 0$.*

**Lemma 1.** *If* (14) *holds for all* $z \in [0,1]^d$ *then*

$$D^{*d}_{N-d+1}\big((u_{N,1},\ldots,u_{N,d}),\ldots,(u_{N,N-d+1},\ldots,u_{N,N})\big) \to 0,$$

*as* $N \to \infty$. *If* (15) *holds for all* $z \in [0,1]^d$ *then*

$$\Pr\Big(D^{*d}_{N-d+1}\big((u_{N,1},\ldots,u_{N,d}),\ldots,(u_{N,N-d+1},\ldots,u_{N,N})\big) > \epsilon\Big) \to 0.$$

*If $u_{N,i}$ are pointwise CUD then they are CUD. If $u_{N,i}$ are pointwise WCUD then they are WCUD.*

*Proof.* The final two statements follow from the first two which we prove here. Pick $\epsilon > 0$ and then choose a positive integer $M > 1/\epsilon$. Next let $\mathcal{L}$ be the lattice of points in $[0,1]^d$ whose coordinates are integer multiples of $1/(2dM)$.

For any $z \in [0,1]^d$ we may choose $z', z'' \in \mathcal{L}$ such that the following hold componentwise: $z' \le z \le z''$, $|z - z'| < \epsilon/(2d)$, and $|z - z''| < \epsilon/(2d)$. Then $0 \le \text{Vol}([0,z'']) - \text{Vol}([0,z]) < \epsilon/2$.

Next, for $N \ge d$, let $\widehat{\text{Vol}}([0,z])$ denote the fraction of the $N-d+1$ points $(u_{N,i},\ldots,u_{N,i+d-1})$ that are in $[0,z]$. Then

$$\widehat{\text{Vol}}([0,z]) - \text{Vol}([0,z]) \le \widehat{\text{Vol}}([0,z'']) - \text{Vol}([0,z]) < \epsilon/2 + \delta^d_{N-d+1}(z''),$$

and with a similarly obtained lower bound via $z'$, we get $D^{*d}_n < \epsilon/2 + \max_{y \in \mathcal{L}} \delta^d_{N-d+1}(y)$. The CUD case follows via (14) with $\epsilon \to 0$. Taking threshold $\epsilon/2$ in (15) yields $\Pr(\max_{y \in \mathcal{L}} \delta^d_{N-d+1}(y) > \epsilon/2) \to 0$, proving the WCUD case. □

The (W)CUD properties are defined in terms of consecutive blocks of observations that overlap, at least when $d > 1$. It is often useful to consider non-overlapping blocks of points such as $(u_1,\ldots,u_d), (u_{d+1},\ldots,u_{2d}),\ldots$. For infinite deterministic sequences it is known (Knuth, 1998, page 155) that if the discrepancy of overlapping sequences tends to zero for all $d$ then the discrepancy of non-overlapping sequences also tends to zero for all $d$. The converse (with 'for all $d$' in both clauses) also holds. See Chentsov (1967).

We need sufficiency of non-overlapping block results for the random case as well. Also it is helpful to be able to work with only a convenient subset of dimensions $d$. Theorem 3 below shows that such special cases are sufficient to prove the WCUD property and hence consistency.

**Theorem 3.** *Let $\mathcal{N}$ be an infinite set of nonnegative integer sample sizes and let $\mathcal{D}$ be an infinite set of nonnegative integer dimensions. Let $u_{N,i}$ be a triangular array for $i = 1,\ldots,N$ and $N \in \mathcal{N}$. For integer $\widetilde{d} \ge 1$ define the nonoverlapping $\widetilde{d}$-tuples $\widetilde{x}^{(i)} = \widetilde{x}^{(i)}(\widetilde{d},N) = (u_{N,\widetilde{d}(i-1)+1},\ldots,u_{N,\widetilde{d}i})$ for $i = 1,\ldots,M = M(N,\widetilde{d}) = \lfloor N/\widetilde{d} \rfloor$. For integer $d$ define the ordinary d-tuples $x^{(i)} = x^{(i)}(d,N) = (u_{N,i},\ldots,u_{N,i+d-1})$ for $i = 1,\ldots,N - d + 1$.*

*Suppose that*

$$\lim_{N \to \infty} \Pr\bigl(D^{*\widetilde{d}}_M(\widetilde{x}^{(1)}, \ldots, \widetilde{x}^{(M)}) > \epsilon\bigr) = 0 \tag{16}$$

*holds for all* $\widetilde{d} \in \mathcal{D}$ *and all* $\epsilon > 0$ *where* $M = M(N, \widetilde{d})$. *Then*

$$\lim_{N \to \infty} \Pr\bigl(D^{*d}_{N-d+1}(x^{(1)}, \ldots, x^{(N-d+1)}) > \epsilon\bigr) = 0 \tag{17}$$

*holds for all* $\epsilon > 0$ *and all integers* $d \geq 1$, *so that* $u_{N,i}$ *are WCUD.*

*Proof.* Let $\epsilon > 0$ and $\eta > 0$, let $d$ be a positive integer, and suppose that $z \in [0,1]^d$. Choose $\widetilde{d} \in \mathcal{D}$ with $d/\widetilde{d} < \epsilon/3$. Because $N$ is tending to infinity, we may assume that $\widetilde{d}/N < \epsilon/3$.

For $i = 1, \ldots, N-d+1$ let $x^{(i)} = (u_{N,i}, \ldots, u_{N,i+d-1})$ and for $k = 1, \ldots, M = \lfloor N/\widetilde{d} \rfloor$ let $\widetilde{x}^{(k)} = (u_{N,\widetilde{d}(k-1)+1}, \ldots, u_{N,\widetilde{d}k})$. Most of the $x^{(i)}$ are nested within exactly one of the $\widetilde{x}^{(k)}$ as follows. For $i = 1, \ldots, N-d+1$, define $k = k(i)$ by $(k-1)\widetilde{d}+1 \leq i \leq k\widetilde{d}$ and define $\ell(i)$ by $\ell = i - (k-1)\widetilde{d}$. If $k(i) \leq M(N,\widetilde{d})$ and $1 \leq \ell \leq \widetilde{d}-d+1$ then the components $u_{N,i}, \ldots, u_{N,i+d-1}$ of $x^{(i)}$ are in positions $\ell$ through $\ell+d-1$ of $\widetilde{x}^{(k)}$. That is $x^{(i)} = \widetilde{x}^{(k)}_{\ell:(\ell+d-1)}$.

Now

$$\sum_{i=1}^{N-d+1} 1_{x^{(i)} \in [0,z]} \leq \sum_{k=1}^{M} \sum_{\ell=1}^{\widetilde{d}-d+1} 1_{x^{((k-1)\widetilde{d}+\ell)} \in [0,z]} + Md + \widetilde{d} - 1$$

$$\leq \sum_{k=1}^{M} f(\widetilde{x}_k) + 2\epsilon N/3,$$

where $f(\widetilde{x}) = \sum_{\ell=1}^{\widetilde{d}-d+1} 1_{\widetilde{x}_{\ell:(\ell+d-1)} \in [0,z]}$. The integral of $f$ over $[0,1]^{\widetilde{d}}$ is $(\widetilde{d}-d+1)\mathrm{Vol}([0,z])$. The function $f$ is piecewise constant within a finite set of axis parallel hyperrectangular regions in $[0,1]^{\widetilde{d}}$. It follows that for some $K < \infty$ $|M^{-1}\sum_{k=1}^{M} f(\widetilde{x}_k) - (\widetilde{d}-d+1)\mathrm{Vol}([0,z])| < K D^{*\widetilde{d}}_M(\widetilde{x}^{(1)}, \ldots, \widetilde{x}^{(M)})$.

Therefore for small enough $\widetilde{\epsilon}_+ > 0$ having $D^{*\widetilde{d}}_M < \widetilde{\epsilon}_+$ will imply that $(N-d+1)^{-1}\sum_{i=1}^{N-d+1} 1_{x^{(i)} \in [0,z]} < \mathrm{Vol}([0,z]) + \epsilon$. Similarly for small enough $\widetilde{\epsilon}_- > 0$ having $D^{*\widetilde{d}}_M < \widetilde{\epsilon}_-$ will imply that $(N-d+1)^{-1}\sum_{i=1}^{N-d+1} 1_{x^{(i)} \in [0,z]} > \mathrm{Vol}([0,z]) - \epsilon$. Therefore when $D^{*\widetilde{d}}_M < \widetilde{\epsilon} = \min(\widetilde{\epsilon}_+, \widetilde{\epsilon}_-)$ we have $\delta^d_{N-d+1}(z) < \epsilon$. By equation (17) we can choose $N \in \mathcal{N}$ large enough that $\Pr(\delta^d_{N-d+1}(z) > \epsilon) < \eta$. Because $z$, $\epsilon$, and $\eta$ are arbitrary we have shown that $u_{N,i}$ are pointwise weakly CUD. To complete the proof we apply Lemma 1. □

## 4. Lattice constructions

Niederreiter (1977) gives a result that shows how lattice rules may be used to construct a triangular array that is CUD. Let $N$ be a prime number. Let $u_0 = 1/N$ and for $i \geq 1$ let $u_i = au_{i-1}/N \mod 1$ where $a$ is a primitive root modulo $N$. For integer dimensions $s \geq 1$, there are $N-1$ distinct consecutive $s$-tuples in this sequence; call them $x^{(i)} = (u_i, \ldots, u_{i+s-1})$. Niederreiter (1977) shows that for well chosen $a = a(N)$ that

$$D_{N-1}^{*s}(x^{(1)}, \ldots, x^{(N-1)}) < \frac{1}{N-1}\left(1 + \frac{(N-2)(s-1)}{\phi(N-1)}\right)\left(\frac{2}{\pi}\log(N) + \frac{7}{5}\right)^s \tag{18}$$

holds, where $\phi$ is Euler's totient function.

The totient function $\phi(n)$ counts the number of positive integers less than or equal to $n$ that are relatively prime to $n$. The totient function grows quickly enough for our purposes because

$$\liminf_{n\to\infty} \frac{\phi(n)}{n}\log(\log(n)) = \exp(-\gamma),$$

where $\gamma \doteq 0.5772$ is the Euler-Mascheroni constant. As a result there exists an $A < \infty$ and an $N_0 < \infty$ such that

$$D_{N-1}^{*s}(x^{(1)}, \ldots, x^{(N-1)}) < \frac{As}{N}\log(\log(N))(\log N)^s \tag{19}$$

holds uniformly in $s \geq 1$ and $N \geq N_0$. The constant $A$ in (19) does not have to grow exponentially with $s$ because the factor $2/\pi$ in equation (18) is smaller than 1.

**Theorem 4.** *Let $\mathcal{N}$ be an infinite set of prime numbers. Let $s(N)$ be a nondecreasing integer function of $N \in \mathcal{N}$ satisfying $s(N) = o([\log(N)/\log(\log(N))]^\alpha)$ for some positive $\alpha < 1$. For each $N \in \mathcal{N}$ let $a(N)$ be a primitive root modulo $N$ for which (18) holds. Form a triangular array via $u_{N,1} = a(N)/N \mod 1$ and $u_{N,i} = au_{N,i-1}/N \mod 1$ for $i = 2, \ldots, N-1$. Then the triangular array $(u_{N,i})$ is CUD.*

*Proof.* For any $d \geq 1$ choose $N_d$ with $s(N_d) \geq d$. Then for all $N \geq N_d$ we have $D_{N-1}^{*d}$ smaller than the right side of (19). Now the growth condition on $s$ makes $D_{N-1}^{*d} \to 0$. □

Owen and Tribble (2005) employed a method of running through the lattice rule more than once, so as to use all $N-1$ of the $s$-tuples in it exactly once. That work also prepends $s$ zeros to the sequence. The result is that $n(N) = N$, and the set of $s$-tuples used to drive the MCMC form a lattice rule Sloan and Joe (1994) in $[0,1]^s$. The lattice rule structure is much more balanced than random $u_i$ would be and this accounts for most of the improved accuracy seen there. Prepending one single $s$-tuple will not affect the CUD property of overlapping $d$-tuples for any $d \geq 1$. Similarly, shifting the points from $u_1, \ldots, u_N$ to $u_{k+1}, \ldots, u_N, u_1, \ldots, u_{k-1}$ for any finite $k$ does not destroy the CUD property (Chentsov (1967)). Finally concatenating a finite number of CUD triangular arrays with the same $N \to \infty$ yields a CUD result.

Lattice rules are commonly randomized via a rotation due to Cranley and Patterson (1976). Let $a \in [0,1]^s$ and suppose that $U \sim U[0,1]^s$. Then the Cranley-Patterson rotation of $a$ is the point $x = a + U \mod 1$ under componentwise arithmetic. Whatever value $a$ has, the point $x$ is uniformly distributed on $[0,1]^s$. A Cranley-Patterson rotation applied to all points in a $s$-dimensional low discrepancy lattice yields a shifted lattice with points that are individually $U[0,1]^s$ while collectively having low discrepancy.

In terms of Figure 1, this proposal starts with the matrix $X$, skipping the randomization of $A$. The matrix $X$ has 0s in its first row and then all other $s$-tuples of the lattice obtained in order with possibly multiple passes. Then a Cranley-Patterson rotation is applied to each row of the matrix. Finally, the concatenation into a single $u$ vector works as in Figure 1.

Owen and Tribble (2005) applied a single Cranley-Patterson rotation to all of the $s$-tuples $(u_{rs-s+1}, \ldots, u_{rs})$ $(r = 1, \ldots, N)$ in the MCMC driving sequence. As we show next, applying a Cranley-Patterson rotation to any CUD or WCUD triangular array yields a triangular array that is WCUD.

**Lemma 2.** *Let $u_{N,i} \in [0,1]$ for $i = 1, \ldots, N$ and $N$ in an infinite set $\mathcal{N}$ of nonnegative integers. Define $v_{N,i} = u_{N,i} + U_{j(i)} \mod 1$ where $j(i) = 1 + (i - 1 \mod m)$, for integer $m \geq 1$. If $u_{N,i}$ are (W)CUD and $(U_1, \ldots, U_m) \sim U[0,1]^m$ independently of $u_{N,i}$, then $v_{N,i}$ are WCUD.*

*Proof.* Suppose that $u_{N,i}$ are WCUD, and let $z \in [0,1]^d$ where $d = rm$ for integer $r \geq 1$. Let $z^{(i)} = (v_{N,di-d+1}, \ldots, v_{N,di}) \in [0,1]^d$ be the $i$'th $d$-dimensional point taken from the rotated triangular array $(v_{N,i})$. Let $x^{(i)} = (u_{N,di-d+1}, \ldots, u_{N,di})$ be the pre-image of $z^{(i)}$ before Cranley-Patterson rotation was applied. Then $z^{(i)} \in [0, z]$ if and only if $x^{(i)} \in B$ where $B = B(z, U)$ is the union of up to $2^d$ axis parallel rectangular boxes in $[0,1]^d$. Therefore the local discrepancy satisfies $\delta^d_{N-d+1}(z) < K D^{*d}_{N-d+1}(x^{(1)}, \ldots, x^{(N-d+1)})$ for some $K < \infty$. It follows that $\Pr(\delta^d_{N-d+1}(z) > \epsilon) \to 0$ for any $z$ and any $\epsilon > 0$. Therefore $\Pr(D^{*d} > \epsilon) \to 0$ for any $d$ that is a multiple of $m$. Therefore $v_{N,i}$ is WCUD. If the $u_{N,i}$ are CUD they are also WCUD and so then $v_{N,i}$ are still WCUD. □

## 5. Consistency of Liao's proposal

Let $a^{(1)}, \ldots, a^{(n)}$ be points in $[0,1]^s$. In Liao's proposal these points are of low $s$ dimensional discrepancy. Let $x^{(i)} = a^{(\tau(i))}$ where $\tau$ is a uniform random permutation of $1, \ldots, n$. Liao's proposal is to concatenate the points $x^{(i)}$ into a driving sequence for MCMC. We need to show that those points are WCUD. It is very natural to make the dimension $s$ of the reordered QMC points equal the number $m$ of driving points consumed by one transition of the chain. It is not however necessary to use $m = s$ as we show below.

Let $u_1, \ldots, u_{sn} \in [0,1]$ be the components of the points $x^{(i)}$ concatenated. The point $u_i$ comes from the $\lceil i/s \rceil$'th point in the sequence, specifically

$$u_i = x^{(\lceil i/s \rceil)}_{i - s(\lceil i/s \rceil - 1)}.$$

TABLE 1
*A sequence of 4 dimensional points $x^{(i)}$, represented in the top row, is concatenated into a sequence of scalars $u_i$, represented in the middle row. Those scalars are then regrouped into the 7 dimensional points $z^{(i)}$ as shown in the bottom row.*

| $x^{(1)}$ | | | | $x^{(2)}$ | | | $x^{(3)}$ | | | | $\cdots$ | | |
|---|---|---|---|---|---|---|---|---|---|---|---|---|---|
| $u_1$ | $u_2$ | $u_3$ | $u_4$ | $u_5$ | $u_6$ | $u_7$ | $u_8$ | $u_9$ | $u_{10}$ | $u_{11}$ | $u_{12}$ | $u_{13}$ | $u_{14}$ | $\cdots$ |
| $z^{(1)}$ | | | | | | | $z^{(2)}$ | | | | | | $\cdots$ |

Now let $z^{(1)}, \ldots, z^{(m)}$ be the points $u_i$ regrouped into $m = \lfloor sn/d \rfloor$ consecutive batches of length $d$, leaving out the last $sn - dm$ of the $u_i$ if $m < sn/d$. That is

$$z_j^{(i)} = u_{d(i-1)+j} = x_{d(i-1)+j-s(\lceil (d(i-1)+j)/s \rceil -1)}^{(\lceil (d(i-1)+j)/s \rceil)}. \tag{20}$$

The situation is illustrated in Table 1 for the case with $s = 4$ dimensional points $x^{(i)}$ regrouped into $d = 7$ dimensional points $z^{(i)}$. The middle row in Table 1 corresponds to the last row in Figure 1 and the bottom row of Table 1 shows the vectors we need to analyze the $d$ dimensional discrepancy of the driving sequence for $d \ne s$.

We will avoid working directly with the rightmost expression in (20) by breaking the $x^{(i)}$ into chunks. To illustrate, consider the point $z^{(2)}$ in the example of Table 1 and let $z \in [0,1]^7$. Then $z^{(2)} \in [0,z]$ if and only if

$$x_1^{(2)} \in [0, z_1], \quad x^{(3)} \in [0, z_{2:5}] \quad \text{and,} \quad x_{1:2}^{(4)} \in [0, z_{6:7}] \tag{21}$$

all hold. The next Lemma takes care of the main details needed to get a bound for the discrepancy.

**Lemma 3.** *For $i = 1, \ldots, n$, let $a^{(i)}$ be points in $[0,1]^s$ with star discrepancy at most $D$. Let $x^{(i)} = a^{(\tau(i))}$ where $\tau$ is a uniformly distributed random permutation of $1, \ldots, n$. For $i = 1, \ldots, \lfloor ns/d \rfloor$ let $z^{(i)}$ be obtained as in equation (20). Assume that $D < 1/3$ and that $n > 3\lceil (d-1)/s \rceil$. Then for any $z \in [0,1]^d$*

$$|\Pr(z^{(i)} \in [0, z]) - Vol([0,z])| \le \frac{3}{2}\Big(2^{1+\lceil (d-1)/s \rceil} - 1\Big)\Big(D + \frac{1}{n}\Big\lceil \frac{d-1}{s} \Big\rceil\Big). \tag{22}$$

*Proof.* Because the $x^{(i)}$ are a permuted version of $a^{(i)}$ they have the same star discrepancy, which is at most $D$. Furthermore for $1 \le j \le k \le s$ the $k - j + 1$–dimensional star discrepancy of $x_{j:k}^{(1)}, \ldots, x_{j:k}^{(n)}$ is at most $D$.

The point $z^{(i)}$ is comprised of chunks taken from consecutive $x^{(j)}$'s. The number of contributing chunks $C$ is between $\lceil d/s \rceil$ and $1 + \lceil (d-1)/s \rceil$ inclusive. The upper limit is attained if the first component of $z^{(i)}$ is the last component of one of the $x^{(j)}$ and the lower limit is attained if the first component of $z^{(i)}$ is the first component of one of the $x^{(j)}$.

For chunks $c = 1, \ldots, C$ let $L(c)$ and $U(c)$ be the first and last indices in $z^{(i)}$ that are taken from chunk $c$. Let $j(i)$ be the integer such that the first component $z_1^{(i)}$ is taken from $x^{(j)}$. Then chunk $c$ of $z^{(i)}$ is taken from $x^{(j(i)+c-1)}$ for $c = 1, \ldots, C$. Let $\ell(c)$ and $u(c)$ be the first and last indices in $x^{(j(i)+c-1)}$ that are used to form the $c$'th chunk of $z^{(i)}$. That is

$$z^{(i)}_{L(c):U(c)} = x^{(j(i)+c-1)}_{\ell(c):u(c)}, \quad c = 1, \ldots, C.$$

Next for $1 \leq \ell \leq u \leq s$ let $N_{\ell:u}(z) = \sum_{i=1}^{n} 1_{x^{(i)}_{\ell:u} \in [0, z_{\ell:u}]}$ count the number of $x^{(i)}$ whose $\ell:u$ subcomponents are in the box $[0, z_{\ell:u}]$.

To streamline notation we write $\widetilde{x}_c$ for the $c$'th chunk $x^{(j(i)+c'-1)}_{\ell(c):u(c)}$, $\widetilde{B}_c$ for the $c$'th box $[0, z_{L(c):U(c)}]$, $\widetilde{v}_c$ for $\mathrm{Vol}(\widetilde{B}_c)$, and $\widetilde{N}_c$ for $N_{\ell(c):u(c)}(z)$. From the discrepancy bounds we know that

$$n(v_c - D) \leq \widetilde{N}_c \leq n(v_c + D).$$

Now $\Pr(z^{(i)} \in [0, z])$ is the product of $C$ conditional inclusion probabilities:

$$\prod_{c=1}^{C} \Pr\Big(\widetilde{x}_c \in \widetilde{B}_c \mid \widetilde{x}_{c'} \in \widetilde{B}_{c'}, \quad 1 \leq c' < c\Big). \tag{23}$$

What is random in (23) is the selection, by simple random sampling, of which $a^{(k)}$ will become $x^{(j(i)+c-1)}$. The conditional probability for chunk $c$ is between $\max((\widetilde{N}_c - c + 1)/(n - c + 1), 0)$ and $\min(\widetilde{N}_c/(n - c + 1), 1)$ depending on how many of the suitable $a^{(k)}$ were 'used up' for chunks $1, \ldots, c - 1$. Clipping these bounds to 0 and 1 is only necessary to handle some extreme cases. We use the notation $y_+$ to denote $\max(y, 0)$.

For the lower bound,

$$\begin{aligned}
\Pr\big(z^{(i)} \in [0, z]\big) &\geq \prod_{c=1}^{C} \frac{(\widetilde{N}_c - C + 1)_+}{n} \\
&\geq \prod_{c=1}^{C} \left(v_c - D - \frac{C-1}{n}\right)_+ \\
&\geq \mathrm{Vol}([0, z]) - (2^C - 1)\left(D + \frac{C-1}{n}\right). \tag{24}
\end{aligned}$$

The last step in (24) is quite conservative. It follows from an expansion of the previous line into $2^C$ terms of which one is $\mathrm{Vol}([0, z])$ and the others have alternating signs and are all of smaller magnitude than $D + (C-1)/n$. The upper bound on $D$ and lower bound on $n$ suffice to give $D + (C-1)/n < 1$ so that the largest terms are not $(D + (C-1)/n))^C$. When $D + (C-1)/n$ is very small then the quantity $2^C - 1$ can be replaced by one almost as small as $C$.

For the upper bound,

$$\Pr(z^{(i)} \in [0, z]) \leq \prod_{c=1}^{C} \frac{\widetilde{N}_c}{n - C + 1} \leq \prod_{c=1}^{C} \frac{n(v_c + D)}{n - C + 1}$$

$$= \prod_{c=1}^{C} \left[ v_c + \left( \frac{v_c + D}{1 - (C-1)/n} - v_c \right) \right]$$

$$\leq \mathrm{Vol}([0, z]) + (2^C - 1) \max_{c \in 1:C} \left( \frac{v_c + D}{1 - (C-1)/n} - v_c \right)$$

$$\leq \mathrm{Vol}([0, z]) + \frac{3}{2}(2^C - 1)\left(D + \frac{C-1}{n}\right). \tag{25}$$

The result follows by combining (24) and (25), and using the fact that $C \leq 1 + \lceil (d-1)/s \rceil$. □

**Theorem 5.** *For $i = 1, \ldots, n$, let $a^{(i)}$ be points in $[0,1]^s$ with star discrepancy at most $D_n^*$. Let $x^{(i)} = a^{(\tau(i))}$ where $\tau$ is a uniformly distributed random permutation of $1, \ldots, n$. For $i = 1, \ldots, \lfloor ns/d \rfloor \equiv \widetilde{n}$ let $z^{(i)}$ be obtained as in equation (20). Suppose that $D_n^* \to 0$ as $n \to \infty$. Then for any $z \in [0,1]^d$ and $d \geq 1$,*

$$E(\delta_{\widetilde{n}}^d(z; z^{(1)}, \ldots, z^{(\widetilde{n})})^2) = O(n^{-1} + D_n^*) \tag{26}$$

*as $n \to \infty$, so that for any $\epsilon > 0$*

$$\Pr(\delta_{\widetilde{n}}^d(z; z^{(1)}, \ldots, z^{(\widetilde{n})}) \geq \epsilon) = O(n^{-1} + D_n^*) \tag{27}$$

*as $n \to \infty$. When $d = 1$, we have the sharper result*

$$|\delta_{\widetilde{n}}^1(z; z^{(1)}, \ldots, z^{(\widetilde{n})})| \leq D_n^*. \tag{28}$$

*Proof.* Let $Y_i = 1$ if $z^{(i)} \in [0, z]$ and $Y_i = 0$ otherwise and let $v = \mathrm{Vol}([0, z])$. Then

$$\delta_{\widetilde{n}}^d(z)^2 = \frac{1}{\widetilde{n}^2} \sum_{i=1}^{\widetilde{n}} \sum_{j=1}^{\widetilde{n}} (Y_i - v)(Y_j - v).$$

For large enough $n$ both $D_n^* < 1/3$ and $n \geq 3d$ hold. Using Lemma 3 we find that $|E(Y_i) - v| \leq K_{11} D_n^* + K_{12}/n$ holds for large enough $n$, where $K_{1\ell} < \infty$ are constants from Lemma 3. Also, $K_{12} = 0$ when $d = 1$.

Now suppose that $Y_i$ and $Y_j$ are well separated in that $|i - j| \geq S \equiv 1 + \lceil (d-1)/s \rceil$. Then none of the points $x^{(k)}$ contribute chunks to both $z^{(i)}$ and $z^{(j)}$. Then the same argument used in Lemma 3 can be adapted to show that for large enough $n$,

$$|\Pr(Y_i Y_j = 1) - v^2| \leq K_{21} D_n^* + K_{22}/n$$

holds for $K_{2\ell} < \infty$. The argument simply requires studying the combined set of chunks for $z^{(i)}$ and $z^{(j)}$. Then $\widetilde{n}^2 E(\delta_n^d(z)^2)$ may be expanded and bounded as follows:

$$E\bigg(\sum_{i=1}^{\widetilde{n}}\sum_{j=1}^{\widetilde{n}} Y_i Y_j - vY_i - vY_j + v^2\bigg)$$
$$\leq \widetilde{n}(2S-1) + \widetilde{n}^2(v^2 + K_{21}n^{-1} + K_{22}D_n^*) - 2\widetilde{n}^2 v(v - K_{11}n^{-1} - K_{12}D_n^*) + \widetilde{n}^2 v^2$$
$$= \widetilde{n}(2S - 1 + K_{21}s/d + 2K_{11}s/d) + \widetilde{n}^2 D_n^*(K_{22} + 2K_{21}).$$

Therefore $E(\delta_n^d(z)^2) = O(n^{-1} + D_n^*)$ as $n \to \infty$, establishing (26) and hence also (27) by Markov's inequality. Finally (28) follows by counting the number of $z^{(i)} \leq z$. □

**Corollary 1.** *Let $a^{(1)}, \ldots, a^{(n)}$ be points in $[0,1]^s$ with star discrepancy $D_n^{*s} \to 0$ as $n \to \infty$. Then the proposal of Liao (1998) is weakly consistent for Metropolis-Hastings sampling.*

*Proof.* Liao's proposal generates pointwise WCUD points by Theorem 5 and hence WCUD points by Lemma 1. They are then weakly consistent by Theorem 1. □

The proposal of Liao (1998) leads to local discrepancies $\delta_n^d(z)$ that vanish, but are not particularly small except for $d = 1$. Liao's motivating application was Gibbs sampling where the number $m$ of variates required for one cycle matches the dimension $s$ of the quasi-Monte Carlo points.

That proposal gets better than Monte Carlo stratification for the $u_i$ individually and for consecutive $q$–tuples such as $(u_{km+r_1}, u_{km+r_2}, \ldots, u_{km+r_q})$ for $k \geq 0$ that nest within consecutive $m = s$–tuples. The proposal does not get particularly good discrepancy even for consecutive pairs $(u_i, u_{i+1})$ because there is a jump from the boundaries of the underlying QMC points when $i$ is a multiple of $s$.

Suppose that we have a problem for which we want to stratify successive updates to the $j$'th component of $\omega$. We might want roughly the right number of low and high proposals for that component and roughly the right probability for consecutive pairs or triples of proposals. In such a case, we might wish to arrange that $s$ consecutive proposals for the $j$'th component of $\omega$ are generated from one of the original $s$ dimensional QMC points.

Similarly we might have a problem in which we wish to treat the acceptance-rejection step of Metropolis-Hastings specially. We could then take $s = m - 1$ and use one QMC point for each proposal we need and one QMC point for each of $s$ consecutive acceptance-rejections.

Whether such alternative schemes work well depends of course on how well the scheme matches the problem. But such schemes can be used to generate consistent samples. Each scheme takes the points of a driving sequence, such as Liao's proposal, groups them into consecutive blocks of $r = ms$ points, and applies a fixed permutation within each block. That permutation operation preserves the WCUD property:

**Theorem 6.** *Let $a^{(i)} \in [0,1]^s$ for $i = 1, \ldots, n$ have star discrepancy $D_n^{*s} \to 0$ as $n \to \infty$. Let $x^{(i)} = a^{(\tau(i))}$ where $\tau$ is a random permutation of $\{1, \ldots, n\}$. Let $v_i = x_{i-s(\lceil i/s \rceil - 1)}^{(\lceil i/s \rceil)}$ be the sequence of x–components for $i = 1, \ldots, ns$.*

*For $r > 1$ let $\sigma$ be an arbitrary permutation of the integers $0, \ldots, r-1$. For $i \geq 1$ let $u_i = v_{j(i)}$ where*

$$j(i) = r\lfloor (i-1)/r \rfloor + \sigma(i - r\lfloor (i-1)/r \rfloor)$$

*Then $u_i$ are WCUD.*

*Proof.* The $v_i$ are the driving sequence proposed by Liao (1998). The $u_i$ are a permutation of them in which no element is moved by more than $r$ positions. The arguments in Lemma 3 and Theorem 5 go through as before. All that has changed is the number and identity of chunks contributing to a consecutive $d$–tuple of $u_i$'s. □

## 6. Acceptance-rejection sampling

It is often impractical, if not impossible, to generate a transition using an *a priori* fixed number of members $u_i$ of the driving sequence. The primary example of such a method is acceptance-rejection sampling, which we sketch here to fix ideas.

To sample a real valued $y$ from a probability mass or density function $f$ we begin by sampling $y$ from $g$ instead where $f(y) \leq cg(y)$ holds for all $y \in \mathbb{R}$ for some constant $c \in [1, \infty)$. Then we accept $y$ with probability $f(y)/(cg(y))$. If $y$ is not accepted then we keep on sampling from $g$ until a point is accepted.

For an illustration of acceptance-rejection sampling suppose that $g$ has CDF $G$ with an efficiently computable inverse $G^{-1}$. Then to get a sample from $f$ let

$$v_j \sim U[0,1], \quad \text{IID}, \quad j \geq 1$$
$$y_j = G^{-1}(v_{2j-1})$$
$$j^* = \min\left\{j \geq 1 \mid v_{2j} \leq \frac{f(y_j)}{cg(y_j)}\right\}$$

and then deliver $y = y_{j^*}$. To use this method one needs to be able to compute the functions $G^{-1}$ and $f/(cg)$. More elaborate versions use $k \geq 1$ uniformly distributed $v_j$ to produce each proposal.

Liao (1998) proposes to handle acceptance-rejection sampling by using the QMC points to make the first two draws from $g$ and the first two acceptance-rejection decisions. If the first two points are both rejected then he suggests drawing from an IID $U[0,1]$ sequence until a point is accepted before switching back to the QMC points. To simplify matters we'll suppose that only one acceptance-rejection step is tried with the QMC points, and that we switch to IID points if that one is rejected. Derandomized adaptive rejection sampling Hörmann et al. (2004) can be used to construct proposals that are accepted with probability arbitrarily close to unity, under a generalized concavity assumption on the density.

To continue the illustration, suppose that the proposal $\widetilde{\omega}$ has 3 components. The first two are generated by inversion, each using one point from $[0,1]$, while the third is done by acceptance-rejection sampling. Then the driving sequence for the first $n$ transitions can be represented in a tableau as follows:

$$
\begin{array}{ccccccccc}
u_1 & u_2 & u_3 & u_4 & (v_{11} & v_{12} & v_{13} & \cdots & ) & u_5 \\
u_6 & u_7 & u_8 & u_9 & (v_{21} & v_{22} & v_{23} & \cdots & ) & u_{10} \\
\vdots & \vdots & \vdots & \vdots & \vdots & \vdots & \vdots & \vdots & & \vdots \\
u_{5i-4} & u_{5i-3} & u_{5i-2} & u_{5i-1} & (v_{i1} & v_{i2} & v_{i3} & \cdots & ) & u_{in} \\
\vdots & \vdots & \vdots & \vdots & \vdots & \vdots & \vdots & \vdots & & \vdots \\
u_{5n-4} & u_{5n-3} & u_{5n-2} & u_{5n-1} & (v_{n1} & v_{n2} & v_{n3} & \cdots & ) & u_{5n}
\end{array}
$$

The points $u_i$ are WCUD and the points $v_{ij}$ are independent $U[0,1]$. The $i$'th row of the table drives the transition from $\omega^{(i)}$ to $\omega^{(i+1)}$. For the proposal $\widetilde{\omega}^{(i+1)}$, $u_{5i-4}$ generates the first component $\widetilde{\omega}_1^{(i+1)}$, $u_{5i-3}$ generates the second component $\widetilde{\omega}_2^{(i+1)}$, $u_{5i-2}$ proposes the third component $\widetilde{\omega}_3^{(i+1)}$, $u_{5i-1}$ is used to accept or reject the third component and $u_{5i}$ is used to accept or reject the entire proposal $\widetilde{\omega}^{(i+1)}$ as $\omega^{(i+1)}$. In the event that $u_{5i-1}$ leads to rejection of the third component, then the infinite sequence $v_{i1}, v_{i2}, \ldots$ is used to continue acceptance-rejection until a third component is generated for the $i$'th proposal.

The difficulty with acceptance-rejection sampling is that the set of driving points for which a transition from $\omega$ to $\omega'$ is proposed does not have a fixed finite dimension. It is a union of regions whose dimensions depend on the number of proposals rejected during the course of acceptance-rejection sampling. This variable dimension complicates discrepancy based methods for studying the driving sequence.

The tabulation above suggests a coupling argument. We replace the sequence $v_{i1}, v_{i2}, v_{i3}, \ldots$ by a single point $v_i \in [0,1]$. The values $v_i$ are independent $U[0,1]$ random variables such that the random variable finally generated by acceptance-rejection would also have been generated by inversion through $v_i$. If the component is continuously distributed with CDF $H$, then $v_i = H(\widetilde{\omega}_3^{(i+1)})$. For discrete $H$ we let $v_i = H(\widetilde{\omega}_3^{(i+1)}-) + \widetilde{v}_i(H(\widetilde{\omega}_3^{(i+1)}) - H(\widetilde{\omega}_3^{(i+1)}-))$ where $\widetilde{v}_i$ is $U[0,1]$ independent of all other driving variables. That is, using acceptance-rejection on the third component can be coupled with the use of inversion for the third component, based on the following driving sequence:

$$
\begin{array}{cccccc}
u_1 & u_2 & u_3 & u_4 & v_1 & u_5 \\
u_6 & u_7 & u_8 & u_9 & v_2 & u_{10} \\
\vdots & \vdots & \vdots & \vdots & \vdots & \vdots \\
u_{5i-4} & u_{5i-3} & u_{5i-2} & u_{5i-1} & v_i & u_{in} \\
\vdots & \vdots & \vdots & \vdots & \vdots & \vdots \\
u_{5n-4} & u_{5n-3} & u_{5n-2} & u_{5n-1} & v_n & u_{5n}
\end{array}
$$

Liao's padding proposal for acceptance-rejection sampling will work so long as inserting an IID $U[0,1]$ sequence into his permuted points at regular intervals, as illustrated above, preserves the WCUD property. This proposal works more generally. If we insert an IID $U[0,1]$ sequence at regular intervals into a CUD or an independent WCUD sequence, then the result is a WCUD sequence. Inserting IID points increases the length of a finite sequence, so that row $N$ of the original sequence becomes row $\widetilde{N}$ of the new sequence.

**Theorem 7.** *Let $v_{N,i} \in [0,1]$ for $i = 1, \ldots, N$ and $N$ in an infinite set of positive integers $\mathcal{N}$. Let $w_i$ for $i \geq 1$ be IID $U[0,1]$. For integers $m \geq 2$ and $b \in \{0, \ldots, m-1\}$ and $i = 1, \ldots, (m+1)\lfloor N/m \rfloor \equiv \widetilde{N}$, let*

$$
u_{\widetilde{N},i} = \begin{cases} w_{\lceil i/m \rceil} & i \equiv b \mod m, \\ v_{N, i - \lceil (i-b)/m \rceil} & else. \end{cases}
$$

*If $v_{N,i}$ are WCUD and independent of $w_i$, then $u_{\widetilde{N},i}$ are WCUD.*

*Proof.* Let $d = r(m+1)$ for integer $r \geq 1$ and choose $z \in [0,1]^d$. For $k = 1, \ldots, \lfloor N/d \rfloor$, the $d$-tuple $x^{(k)} = (u_{\widetilde{N},(k-1)d+1}, \ldots, u_{\widetilde{N},kd})$ has $r$ components from $w$ and $dr$ components from $v_{N,i}$. Let $A$ represent the components from $w$ and $B$ represent the components from $v_{N,i}$. Then

$$
N(z) \equiv \sum_{i=1}^{\lfloor N/d \rfloor} 1_{z^{(i)} \in [0,z]} = \sum_{i=1}^{\lfloor N/d \rfloor} 1_{z_A^{(i)} \in [0,z_A]} 1_{z_B^{(i)} \in [0,z_B]}.
$$

The $z_B^{(i)}$ are IID Bernoulli variables taking 1 with probability $\text{Vol}([0, z_B])$ independently of $z_A^{(i)}$. Therefore $\text{Var}(N(z) \mid w_1, \ldots) \to 0$ as $N \to \infty$, while $E(N(z) \mid w_1, \ldots) \to \text{Vol}([0, z_B]) \sum_{i=1}^{\lfloor N/d \rfloor} 1_{Z_A^{(i)} \in [0, z_A]}$. Now $\sum_{i=1}^{\lfloor N/d \rfloor} 1_{Z_A^{(i)} \in [0, z_A]}$ converges in probability to $\lfloor N/d \rfloor \text{Vol}([0, z_A])$ as $N \to \infty$ because $v_{N,i}$ are WCUD.

It follows that $\Pr(\delta^d_{\lfloor N/d \rfloor}(z) > 0) \to 0$ as $N \to \infty$. Invoking Lemma 1 and Theorem 3 completes the proof. □

If some fixed number $k \geq 1$ of components are to be sampled by acceptance-rejection at each step, then we simply apply Theorem 7 $k$ times inserting $k$ independent streams of IID $U[0,1]$ random variables.

**Remark 1.** *It is important that one of the streams in Theorem 7 be IID. Merging two independent and WCUD streams does not necessarily produce a WCUD result. For example if the two WCUD sequences $u_i$ and $v_i$ are independent Cranley-Patterson rotations of the same underlying deterministic sequence, then $u_1, v_1, u_2, v_2, \ldots$ will not be WCUD because every pair of the form $(u_i, v_i)$ will lie within one or two lines in $[0,1]^2$.*

## 7. Example: Probit regression

In this section we apply a Gibbs sampling scheme developed by Albert and Chib (1993) for a probit regression example of Finney (1947). For $i = 1, \ldots, 39$, the response $Y_i \in \{0, 1\}$ is 1 if the subject exhibited vasoconstriction and 0 otherwise. The predictors are $X_i = (X_{i1}, X_{i2})$ where the first component is the volume of air inspired and the second is the rate at which air is inspired. The probit model has $Y_i = 1_{Z_i > 0}$ where $Z_i \sim N(\beta_0 + \beta_1 X_{i1} + \beta_2 X_{i2}, 1)$ are conditionally independent given $X_1, \ldots, X_n$ and $\beta = (\beta_0, \beta_1, \beta_2)'$. The data are shown in Figure 2.

Taking a non-informative prior for $\beta$, the full conditional distribution of $\beta$ given $Z_1, \ldots, Z_n$ is $N((X'X)^{-1}X'Z, (X'X)^{-1})$, where $X$ is the $n$ by 3 matrix with $i$'th row $(1, X_{i1}, X_{i2})$ and $Z$ is the column vector of $Z_i$ values. If $Y_i = 1$ then the full conditional distribution of $Z_i$ given $\beta$ and $Z_j$ for $j \neq i$ is $N(\beta_0 + X_{i1}\beta_1 + X_{i2}\beta_2, 1)$ truncated to $[0, \infty)$. If $Y_i = 0$ then full conditional distribution is instead truncated to $(-\infty, 0]$.

To run the Gibbs sampler we need only invert the normal CDF to obtain the normal and truncated normal full conditionals. We used the SSJ package of L'Ecuyer (2008) which includes a high accuracy inverse normal CDF based on Blair et al. (1976).

The dimension of this simulation is $m = 42$ variables per step. In our lattice sampling we let $N$ be a prime number and choose $a$ to be a primitive root modulo $N$. Suppose at first that $m$ and $N - 1$ are relatively prime. Then the driving sequence we use is obtained by scanning the following matrix from left to right and top to bottom,

$$\begin{pmatrix} 0 & 0 & 0 & \cdots & 0 \\ 1 & a & a^2 & \cdots & a^{m-1} \\ a^m & a^{m+1} & a^{m+2} & \cdots & a^{2m-1} \\ a^{2m} & a^{2m+1} & a^{2m+2} & \cdots & a^{3m-1} \\ \vdots & \vdots & \vdots & \ddots & \vdots \\ a^{(N-2)m} & a^{(N-2)m+1} & a^{(N-2)m+2} & \cdots & a^{(N-1)m-1} \end{pmatrix} \mod N$$

and then dividing by $N$ and applying the same Cranley-Patterson rotation to each $m = 42$ dimensional row. Starting in the second row above, the matrix above contains elements from a linear congruential generator (LCG), with initial seed 1. Unlike LCGs, integration lattices contain a point at the origin, introduced here via the first row.

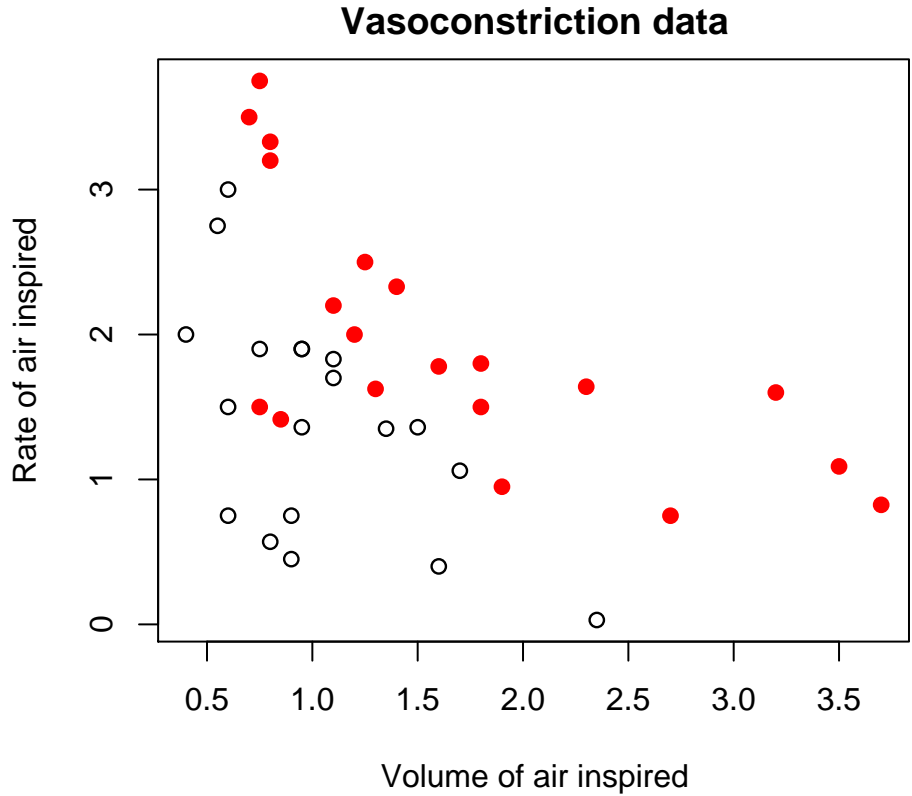

FIG 2. *Vasoconstriction data from Finney (1947) as described in the text. Cases with vasoconstriction are solid points. There are two solid points at* (1.9, 0.95).

When $m$ and $N-1$ are not relatively prime, then the simple scheme above does not utilize all $N-1$ $m$-tuples of the LCG. For the general case let $g = \mathrm{GCD}(m, N-1)$, so $g = 1$ when $m$ and $N-1$ are relatively prime. Then the table above contains the 0 vector and $g$ identical blocks of $b = (N-1)/g$ rows. We multiply the $k$'th such block by $a^{k-1}$ (modulo $N$) in order to get all $m$-tuples of the LCG into the simulation.

Ideally $a$ should be a good multiplier for a linear congruential generator, so that consecutive tuples are nearly equidistributed. If possible $a^m$ should also be a good multiplier so that consecutive updates to a given parameter are also well equidistributed.

We applied our technique with values of $N$ and $a$ shown in Table 2. These are from L'Ecuyer (1999). We also applied Liao's method on the same problem. Each method was repeated independently 300 times in order to estimate the sampling variance. Table 3 shows estimated variance reduction factors for the posterior means of $\beta_j$. The 0.975 point of the $F_{299,299}$ distribution is 1.25. Therefore

TABLE 2
*Shown are the parameters of the LCGs used to drive the Gibbs sampler for the Finney (1947) probit model. Five prime numbers N near powers of two are shown. Five corresponding primitive roots a from L'Ecuyer (1999) are listed. In each case g is the greatest common divisor of a and $N-1$. The simulation goes through g blocks of $b = (N-1)/g$ m-tuples (m = 42) from the LCG.*

| N | 1021 | 2039 | 4093 | 8191 | 16381 |
|---|------|------|------|------|-------|
| a | 65   | 393  | 235  | 884  | 665   |
| g | 6    | 2    | 6    | 42   | 42    |
| b | 170  | 1019 | 682  | 195  | 390   |

TABLE 3
*This table shows variance reduction factors comparing WCUD-MCMC with IID-MCMC. There are 5 sample sizes N and three regression parameters $\beta_j$. We are estimating the posterior mean of $\beta_j$ and comparing variances of these estimates. The upper block compares a Cranley-Patterson rotated LCG to IID sampling. The middle block compares Liao's permutation scheme, on the same rotated LCG, to IID sampling. The bottom block compares LCG sampling to permutations. Individual entries between 0.8 and 1.25 are not statistically significantly different from 1.*

|       |         | N | 1021 | 2039 | 4093 | 8191 | 16381 |
|-------|---------|---|------|------|------|------|-------|
| LCG   | $\beta_0$ |   | 15.9 | 29.9 | 22.5 | 44.4 | 37.6  |
| vs    | $\beta_1$ |   | 14.9 | 29.7 | 23.3 | 41.9 | 39.1  |
| IID   | $\beta_2$ |   | 17.1 | 27.4 | 22.9 | 46.1 | 35.2  |
| Liao  | $\beta_0$ |   | 20.0 | 17.9 | 23.1 | 19.0 | 19.0  |
| vs    | $\beta_1$ |   | 18.5 | 18.5 | 21.7 | 19.8 | 20.2  |
| IID   | $\beta_2$ |   | 21.3 | 16.6 | 24.1 | 20.0 | 18.5  |
| LCG   | $\beta_0$ |   | 0.79 | 1.67 | 0.97 | 2.24 | 1.98  |
| vs    | $\beta_1$ |   | 0.80 | 1.60 | 1.07 | 2.18 | 1.93  |
| Liao  | $\beta_2$ |   | 0.80 | 1.64 | 0.95 | 2.30 | 1.91  |

individual ratios between 0.8 and 1.25 should not be considered statistically significant.

Some trends are clear in Table 3. The variance reductions for all three regression coefficients track each other very closely. Liao's method typically gives a variance reduction of about 20 fold. The LCG method gives a variance reduction that tends to increase with sample size but is not perfectly monotone. The quality of the underlying lattices may not be monotone in $N$. For larger $N$ the LCG approach performed better than Liao's. For smaller $N$, Liao's approach did slightly better but perhaps not significantly so.

The QMC-MCMC methods also reduce the variance of the estimated posterior means for the latent parameters $Z_1, \ldots, Z_{39}$, sometimes by very large amounts. When $|Z_i|$ is large then the variance reduction for it is nearly the same as we see for the coefficients $\beta_j$. When $|Z_i|$ is small, corresponding to cases near the borderline, then variance reductions of several hundred fold are attained.

Figure 3 plots the variance reductions for latent parameters versus the estimated values of those latent parameters. The curves corresponding to the largest and smallest sample sizes are shown. The curves for the other sample sizes are qualitatively similar. The LCG version attains some much larger variance re-ductions, sometimes over 500–fold, for the $Z_i$ near 0. Table 4 shows summary statistics of the variance reductions.

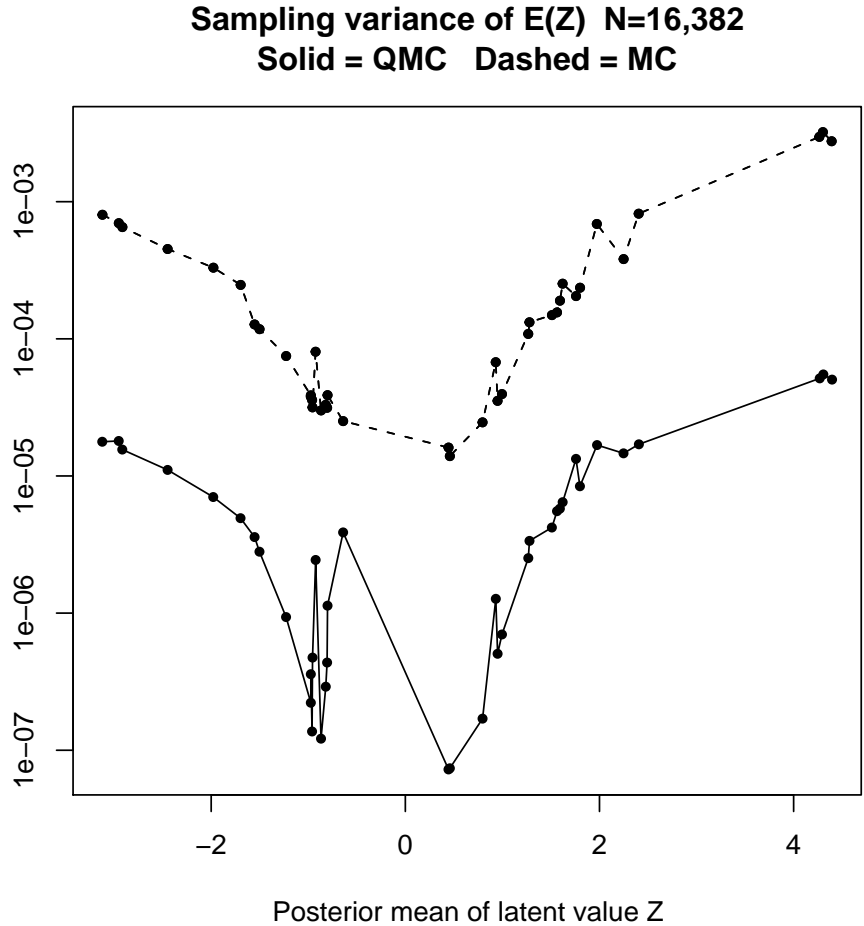

FIG 3. *This figure shows the sampling variance of MCMC estimates of $E(Z_i \mid Y_1, \ldots, Y_{39})$ for the Finney data. The posterior means themselves are on the horizontal axis. Their variance is on the vertical axis with QMC-MCMC values joined with a solid line and IID-MCMC values joined with a dashed line. The displayed data are for $N = 16{,}384$.*

In MCMC sampling there is usually a bias because the chain only approaches its stationary distribution asymptotically. Variance reductions are most meaningful when the biases of two methods are comparable and small. Because the sample values of all 42 parameters averaged over 300 replications are essentially identical for all 3 methods at every $N$, we are sure that the biases of all of these methods are nearly identical here. Table 5 summarizes the evidence on bias.

TABLE 4
*This table summarizes variance reduction factors for the posterior means of the latent parameters $Z_i$, pooling all 5 run lengths $N$. The min, max, mean and quartiles of the variance reduction factors are shown.*

|  | Min | $Q_{0.25}$ | $Q_{0.5}$ | Mean | $Q_{0.75}$ | Max |
|---|---|---|---|---|---|---|
| LCG | 10.0 | 23.5 | 38.9 | 68.1 | 67.9 | 561.6 |
| Liao | 11.1 | 18.6 | 22.5 | 37.5 | 48.4 | 157.0 |

TABLE 5
*This table summarizes parameter differences, averaged over replications, between the sampling methods. All parameters and all sample sizes are included. The first row compares LCG-MCMC to IID-MCMC. The second row compares permuted QMC to IID sampling. The third row compares two versions of QMC-MCMC. Corresponding mean and median values (not shown) are all within the range $\pm 1.5 \times 10^{-4}$.*

|  | Min | $Q_{0.25}$ | $Q_{0.75}$ | Max |
|---|---|---|---|---|
| LCG − MC | −0.0053 | −0.00072 | 0.00053 | 0.0073 |
| Liao − MC | −0.0081 | −0.00068 | 0.00049 | 0.0087 |
| Liao − LCG | −0.0029 | −0.00017 | 0.00022 | 0.0014 |

## 8. Discussion

This paper has produced some specific constructions of WCUD sequences, has given general methods that convert WCUD sequences into other WCUD sequences, and has found conditions that simplify the task of proving that a sequence is WCUD.

Some further results appear in the thesis of Tribble (2007). In particular, Tribble (2007) establishes results parallel to the ones here, for methods that use small feedback shift register generators instead of small congruential generators. Tribble (2007) also introduces a skipping method that simplifies the process of running through all $s$-tuples of a small random number generator.

Our motivating application for studying (W)CUD sequences is for MCMC, especially in continuous state spaces. The sequences we construct take place in a continuous space, and the transformations we apply are those for continuous random variables. For technical reasons, we have analyzed the methods for discrete state spaces. We expect that some further though hopefully mild regularity will be required. For now, it is encouraging that nothing seemed to go awry in the continuous example that we ran.

Our analysis of Liao's shuffling proposal shows that it only improves the rate of convergence for one dimensional discrepancies. This fact suggests that his proposal will affect the constant, but not ordinarily the rate in the MCMC convergence. The numerical results appear to bear this out. It is however surprising to see as much as a 20 fold variance reduction from a method that only improves certain one dimensional histograms of the input sequence. This must mean that those one dimensional aspects are relatively important compared to high dimensional and more subtle features. Such a pheomenon has been seen before in finite dimensional applications of QMC. The effective dimension can be much smaller than the nominal one as described in Caflisch et al. (1997). Of course not all integrands have low effective dimension and we would not expect large variance reductions every time MCMC was applied. Furthermore the posterior moments we studied are smooth functions of their arguments and this plays to a strength of QMC.

The LCG scheme by contrast shows steady improvement with increasing sample size, though we have no theory that applies to the rate of convergence, and no reason to expect that better than the MC rate can be attained for MCMC problems. Against the possibility of better LCG performance there is a tradeoff. Liao's method is very simple to use and LCGs require parameter searches.